# The glass transition diagram in model metallic glasses

X.Q. Gao, W.H. Wang, K. Zhao, H.Y. Bai*

*Institute of Physics, Chinese Academy of Sciences, Beijing 100190, People's Republic of China*

**Abstract**

We report a strain rate (equivalent to experimental observation time) induced glass transition in model SrCaYbMg(Li)Zn(Cu) metallic glasses at room temperature. A critical strain rate, equivalent to glass transition temperature, is found for the strain rate induced a glassy state to liquid-like viscoplastic state translation. The results show that the observation time, equivalent to temperature and stress, is a key parameter for the glass transition. A three-dimension glass transition phase diagram involved in time, temperature and stress in metallic glasses is established for understanding the nature of the metallic glasses.

# I. Introduction

Glasses are regarded as liquids that have lost their flow ability. Generally, it is recognized that the glass transitions between supercooled liquid and glassy states in metallic glasses (MGs) are induced by thermalization or by applied stress [1-6]. Through simulation Guan *et al.*[6] quantified the intimate coupling of temperature and shear stress in glass transition. Liu and Nagel unified the intrinsic parameters of temperature, stress and density to jamming, and provided a jamming diagram to unify the description of jamming in colloidal glasses[7]. The glass transition is observation time dependent[3]. For polymer glasses, their responses to stress can be observed both within a short time at high temperature and a sufficient long time at low temperatures, which is time-temperature superposition principle[8]. The physical similarity among MGs, colloidal and polymer glasses suggests that the glass transition in MGs could have the similar response to the key parameters of temperature ($T$), stress as well as the observation time.

However, little work has been done on the effects of observation time on glass transition in MGs. This is due to the MGs normally have high flow energy barrier below glass transition temperature ($T_g$), and the homogeneous deformation of MGs generally is only observed at high $T$ or in inaccessible long time scale under stress at low $T$ [9-10]. For example, for Zr-based MG with activation energy of flow units ~1.3 eV, the room-temperature (RT) uniaxial compression at 80% of its yield stress for 5 hours only leads to homogeneous flow with irreversible strain of ~2.0x10$^{-4}$.[11] The long time scale and very small homogeneous flow strain at low $T$ make the study of the effects of strain rate and observation time on glass transition very difficult. Recently, Sr-based MGs with low $T_g$ have been developed. Remarkable liquid-like homogeneous flow near room temperature and under relatively high strain rates can be realized in the glasses due to their very low flow activation energy[12-13]. Importantly, the supercooled liquid state of the MGs can be realized solely by the homogeneous mechanical behavior, which provides a model system to study the roles of strain rate or observation time in glass transition, and to test the concept of glass transition phase diagram in MGs.

In this work, the glass to supercooled liquid state transition solely induced by the strain rate $\dot{\gamma}$ in the model SrCaYbMg(Li)Zn(Cu) MG systems. We examine the role of $\dot{\gamma}$, which is equivalent to the observation time, in the glass transition of the MGs, and a time involved glass transition phase diagram is obtained. The results show that the observation time, temperature and stress can be incorporated into a generalized description of glass transition in MGs, and confirm the unifying link between the glass transition and jamming.

**II. Experiments**

Cylindrical specimens of $Sr_{20}Ca_{20}Yb_{20}Mg_{20}Zn_{20}$, $Sr_{20}Ca_{20}Yb_{20}Mg_{20}Zn_{10}Cu_{10}$, $Sr_{20}Ca_{20}Yb_{20}(Li_{0.55}Mg_{0.45})_{20}Zn_{20}$ MGs with 2 mm diameter were prepared by copper mold casting method[13]. The $T_g$ and shear modulus $G$ of the MGs were presented in Table I. Compression tests at RT were conducted on these MGs with a length-to-diameter ratio of 2 at different applied strain rates on Instron 3384 machine (Norwood, MA). The temperature of the compression tests was controlled by thermocouple and the tests were in pure Ar atmosphere to avoid the oxidation. The X-ray diffraction showed that the temperature in the range of 310- 350 K for compression tests did not induce the crystallization. The scanning electron microscopy (SEM) conducted in a Philips XL30 instrument. According to the shear cooperative model[14], the barrier energy of plastic flow units $W_{STZ}$ [also termed as shear transformation zones (STZs)] in MGs can be estimated by[15] $W_{STZ} \propto 0.4GV_m$, where $V_m$ is average molar volume. These MGs have ultra low $G$ (see Table I), and their flow barrier energy $W_{STZ}$ is estimated to be 0.79, 0.75, and 0.52 eV, respectively. Clearly, these MGs have much lower activation energy of flow units compared with that of conventional MGs[15] and the activation of flow units is then much easier. We then could realize the homogenous flow or glass to supercooled liquid transition near RT in the MGs by stress in reasonable observation time scale, and characterize the effect of time on glass transition through their homogeneous mechanical behavior.

## III. Results and discussions

Figure 1 shows the compressive stress–strain curves at different applied strain rates for as-cast rods of $Sr_{20}Ca_{20}Yb_{20}Mg_{20}Zn_{20}$ [Fig. 1(a)] and $Sr_{20}Ca_{20}Yb_{20}(Li_{0.55}Mg_{0.45})_{20}Zn_{20}$ [Fig. 1(b)]. For both of the MGs, at RT and above a critical strain rate $\dot{\gamma}_g$, the true stress–strain curves exhibit an elastic loading region followed by catastrophic failure into small pieces without any plastic deformation [see Fig.1(c)]. The fracture approaches the ideal brittle behavior[16], which indicate that at RT the MG is in glassy state. However, below a critical strain rate $\dot{\gamma}_g$, remarkably, the MGs displays a stress overshoot which is often observed in the supercooled liquid state of MGs at high $T$ (e.g. for Zr-based MGs, the superplastic flow occurs above 600 K [9]). The stress overshoot is due to the structural evolution (i.e. accumulation of free volume) accompanied by softening[16-17]. Due to the intrinsic relaxation, the stress attains a steady state after the overshoot [Fig. 2 (a)-(b)] which is a typical feature of homogeneous flow in glass-forming liquid. As shown in Fig.1(d), the MGs can be deformed to 50% of its original height without observable cracking and shear banding, and further deformation is still possible, implying the significant homogeneous flow of the MG at RT. This means the metallic glasses transform into supercooled liquid state at RT by strain rate, that is the strain rate induced glass transition. To confirm the homogeneous deformation at RT, we examined the surface of the heavily deformed MG to see if there exist nano-scale shear bands. Figure 2 shows the SEM images of the compressed sample with a strain of ε ≈ 50%. Although some oxidation spots appear, the severe deformed specimen almost keeps the smooth surface of the as-cast state and no primary shear bands and other observable shear bands can be found. The results confirm that the liquid-like homogeneous flow, similar to that observed in supercooled liquid in conventional MGs at high $T$, can be realized solely by $\dot{\gamma}$ at RT in the MGs.

Similar to the $T$ dependence of the homogenous deformation in supercooled liquids, a strain rate dependent homogeneous flow can be clearly seen in the MG below $\dot{\gamma}_g$ [see Fig. 1(b)]. When the $\dot{\gamma}$ decreases, the yielding stress and steady flow stress decrease. The $\dot{\gamma}$ dependence of the steady state flow stress ($\sigma_{flow}$) corresponds to the non-Newtonian flow. At relatively low strain rates, the rate of free volume annihilation gradually comes to match that of production, and a steady-state condition is eventually achieved[16-17]. Meanwhile, the overshoot peak stress (or the maximum flow stress $\sigma_{max}$) decreases monotonically from ~400 MPa to ~100 MPa as $\dot{\gamma}$ decreased from $2\times10^{-4}$ to $5\times10^{-6}$ s$^{-1}$ in the homogeneous deformation region, and the peak stress almost disappeared when $\dot{\gamma} = 5\times10^{-6}$ s$^{-1}$. The variation tendency of the steady state stress is similar to that of the overshoot peak stress, which is a typical homogeneous flow behavior in supercooled liquid state. The results clearly illustrate the similar roles of $\dot{\gamma}$ and $T$ on the homogeneous flow [9].

Figure 3 shows the maximum stress as a function of $\dot{\gamma}$ ranging from $5\times10^{-6}$ to $3\times10^{-4}$ Pa·s in Sr$_{20}$Ca$_{20}$Yb$_{20}$(Li$_{0.55}$Mg$_{0.45}$)$_{20}$Zn$_{20}$ and actually illustrates the glassy state (brittle) to liquid-like state (homogeneous flow) translation induced by strain rate. The $\sigma_{max}$ monotonically increases with the increasing of $\dot{\gamma}$ in homogeneous regime, and in glassy state $\dot{\gamma}$ reaches a plateau of maximum stress. A well-defined transition from supercooled liquid state to glassy state can be seen as $\dot{\gamma}$ reaches a critical value of $\dot{\gamma}_g = 3\times10^{-4}$ s$^{-1}$ (as indicated in Fig.3). The critical value $\dot{\gamma}_g$, which is equivalent to $T_g$ in temperature induced glass transition, is the intrinsic parameter characterizing the glass translation induced by strain rate or action time.

The relationship of apparent viscosity $\eta$ of homogeneous flow, $\dot{\gamma}$ and the flow stress $\sigma_{flow}$ can be expressed as [18-19]: $\dot{\gamma} = \sigma_{flow} / 3\eta$. The dynamical definition of glass transition is the viscosity of the supercooled liquid reaches to ~$10^{13}$ Pa [16, 20-21]. We then estimate the critical strain rate $\dot{\gamma}_g^{cal}$ by $\dot{\gamma} = \sigma_{max} / 3\eta$ (The $\sigma_{max}$ is close to yield strength). The estimated values of $\dot{\gamma}_g^{cal}$, as listed in Table 1, are very close to the experimental determined values for these MGs. Meanwhile, from Table I, one can see that the MG with larger $T_g$ has smaller $\dot{\gamma}_g$, which indicates that the $T_g$ correlates well to $\dot{\gamma}_g$ for these MGs. This further confirms that the $\dot{\gamma}$, equivalent to temperature, is another key parameter to describe the glass transition.

Maxwell firstly extrapolated the relation between liquid and solid behavior, and proposed the relation between the viscosity and relaxation time τ as: $\tau=\eta/G_\infty$, where $G_\infty$ is the instantaneous shear modulus. The Maxwell relaxation time τ provides a key to understanding the glass transition [22-23]. The general property of the relation indicates that liquid is solid-like on time scales much shorter than τ, and on a sufficiently long time scale any glass behaves like a liquid. The glass transition is observation time dependent, and difference between glasses and fluid liquid can be defined by a nondimensional number of the so-called Deborah number [22-23]: $D= t_r/t_o$, where $t_r$ is the time of relaxation, and $t_o$ is the time of observation. If $t_o$ is very large, you see the glass flowing. On the other hand, if $t_r \gg t_o$, the liquid is a glassy solid. A viscous liquid can be viewed as solid which flow in enough short time scale, and a glass can be regarded as viscous liquid in enough large time scale [23]. Actually, the inverse strain rate is related to action time scale of strain or stress on MGs, and is equivalent to the experimental observation time, and the effect of the strain rates on the glass transition is equivalent to that of the observation time on glass transition.

To determine the relation between the time scale, temperature and stress in glass transition process of MGs, we studied the relation between maximum yield stress and inverse strain rate at RT. Figure 4(a) shows the $\sigma_{max}$ vs. $\dot{\gamma}$ in the MG. When the liquid-to-glass transition occurs, the $\sigma_{max}$ decreases with the extending observation time, but apparently not in a linear fashion. According to the general steady state flow theory [24], the relation between the σ and $\dot{\gamma}$ is given by [24-25],

$$\dot{\gamma} = 2c_f v_D \exp\left(-\frac{\Delta G^m}{kT}\right) \sinh\left(\frac{\sigma V}{2\sqrt{3}kT}\right) \quad (1)$$

where $v_D$ is the Debye frequency, $\Delta G_m$ is the activation free energy of defect migration,

$V$ is the activation volume and $k$ is Boltzmann constant. At a given temperature of RT, the $v_D$, $\Delta G_m$ and $V$ are constant, and the relation between the $\sigma_{max}$ and $\dot{\gamma}$ is[24-25],

$$\dot{\gamma}^{-1} = \dot{\gamma}_0^{-1} \sinh^{-1}(a\sigma_{max}) \qquad (2)$$

where $\dot{\gamma}_0 = 2c_f v_D \exp\left(-\dfrac{\Delta G^m}{kT}\right)$ and $a = \dfrac{V}{2\sqrt{3}kT}$. The line in Fig. 4(a) fits well the relation between inverse strain rate and $\sigma_{max}$ within the error. This indicates the shape of the phase boundary between stress and reverse strain rate in glass transition diagram should be a concave curvature.

Figure 4(b) shows the relation of $T$ and $\dot{\gamma}_g$ during glass transition for $Sr_{20}Ca_{20}Yb_{20}Mg_{20}Zn_{10}Cu_{10}$ MG. The $\dot{\gamma}_g$ is 5x10$^{-7}$, 1x10$^{-5}$ and 8x10$^{-5}$ s$^{-1}$ at 300, 323 and 343 K, respectively. This indicates that much longer observation time, depending on the system relaxation time, is needed to observe the viscoplastic flow in MG at lower $T$. For the MGs with high $T_g$, the relaxation process is extremely sluggish at RT, and the observation time has to be extended too long time scale for observing the glass transition, the homogeneous flow is inconspicuous or even not be experimentally achieved at RT. This is confirmed in Zr-based MG with higher flow activation energy. The homogeneous viscous flow with irreversible strain at RT by uniaxial compression is observable only when $\dot{\gamma}$ is less than $\sim 10^{-10}$ s$^{-1}$.

The phase diagrams involved in temperature and stress to describe the jamming and glass transition in different glasses are summarized [6,7,26]. Based on previous and our above results, we propose a speculative 3D glass transition phase diagram involved in observation time $t$ (equivalent to $\dot{\gamma}^{-1}$), $T$ and stress $\sigma$. The interdependence of $t$, $T$ and $\sigma$ allows to plot the schematic 3D diagram for MGs as shown in Fig. 5. We use the curve forms of Fig. 4 to determine the phase boundaries. And we note that the values of the coefficient and shape of the glass transition surface may be different for details of specific glass systems [26-27]. The glass region, near the origin, is enclosed by the depicted surface. According to the diagram, glass transition can be realized either by $T$, or by external stress or by changing observation time scale.

We draw a horizontal cross-section through point $t_0$, which represents the usual experimental observation time scale (100 s) of glass transition[28], and the temperature corresponding to point *A* is the usual $T_g$. In the case of no applied stress, in the limit of $T/T_g<<1$, the glass we see will flow if the observation time is sufficient long (or the strain rate is slow enough), which has been confirmed by experiments[10-12] and can interpret our observations in Sr-based MGs. By contrast, the liquid is solid glass-like as well when the observation time is infinite short (equivalent to extreme large strain rate). Actually, in the supercooled liquid region, it is found that increase in strain rate can lead to a transition from homogeneous flow to inhomogeneous deformation, and makes the liquids behave solid like [29-30]. Accordingly, the time, stress and temperature play the similar roles in glass transition, and the three parameters at each corner are divergent. The divergence in time axis reflects that a sufficiently short time scale any viscous liquid behaves like a solid glass and a solid glass can be viewed as viscous liquid when the time scale is extended to infinite. The diagram helps rationalize and unify the description of apparently diverse transitions from liquid-like to solid-like behaviors by means of glass transition in MGs. For practical applications, the diagram provides guidance in the parameters control to achieve a desired behavior and to search new glasses.

### IV. Conclusions

A glass to liquid-like homogeneous flow translation at RT occurs in Sr-based MGs when the strain rate reaches a critical value, which indicates the action time of strain which is equivalent to observation time, like temperature and stress, is another intrinsic parameter to understand the glass transition. A 3D phase diagram involved in the observation time, stress and temperature is established to describe the glass transition in MGs. The temperature, observation time, and stress are equivalent control parameters, and the change of any of them having equivalent role in glass transition. The glass transition is interplay between thermal energy, applied stress and action time scale. The work might be helpful for understanding the nature of the glass transition and for providing a unifying link between the glass transition and jamming.

This work was supported by MOST 973 of China (Nr. 2010CB731603) and the NSF of China (Nrs. 51271195 and 51071170).

# References

[1] P. N. Pusey and W. van Megen, Phys. Rev. Lett. 59, 2083 (1987).

[2] C. A. Angell, and L. V. Woodcock, Adv. Chem. Phys. 48,397 (1981).

[3] E. Leutheusser, Phys. Rev. A 29, 2765 (1984).

[4] W.H. Wang, J Appl. Phys. 110, 053521 (2011).

[5] B. Yang, C. T. Liu, T. G. Nieh, M. L. Morrison, P. K. Liaw, & R. A. Buchanan, J. Mater. Res. 21, 915 (2006).

[6] P. F. Guan, M. W. Chen, & T. Egami, Phys. Rev. Lett. 104, 205701 (2010).

[7] A.J. Liu and S.R. Nagel, Nature 396, 21 (1998).

[8] N. B. Olsen, T. Christensen, and J. C. Dyre, Phys. Rev. Lett. 86, 1271 (2001).

[9] J. Lu, G. Ravichandran, W.L. Johnson, Acta Mater. 51, 3429 (2003).

[10] K.W. Park, J.C. Lee. Acta Mater. 56, 5440 (2008).

[11] H.B. Ke, P. Wen, W. H. Wang, & A. L. Greer, Scripta Mater. 64, 966 (2011).

[12] K. Zhao, X. X. Xia, H. Y. Bai, W.H. Wang, Appl. Phys. Lett. 98, 141913 (2011).

[13] X. Q. Gao, H. Y. Bai, W.H. Wang, J. Non-Cryst. Solids 357, 3557 (2011).

[14] W. L. Johnson, K. Samwer, Phys. Rev. Lett. 95, 195501(2005).

[15] H. B. Yu, *et al.* Phys. Rev. B 81, 220201 (2010).

[16] C. A. Schuh, T. Hufnagel, Acta Mater. 55, 4067 (2007).

[17] A. S. Argon, Acta Metall. 27, 47 (1979).

[18] Y. Kawamura, A. Inoue, and T. Masumoto, Mater. Trans., JIM 40,336 (1999).

[19] F. A. McClintock and S. A. Argon, *Mechanical Behavior of Materials* (Addison-Wesley, Reading, MA, 1966), p. 273.

[20] L. M. Martinez, C. A. Angell, Nature 410, 663 (2011).

[21] M. L. Falk, and J. S. Langer, Ann. Rev. Condensed Matter Phys. 2, 353 (2011).

[22] M. Reiner. Phys. Today 17, 62 (1964).

[23] J. C. Dyre, Rev. Mod. Phys. 78, 953, (2006).

[24] F. Spaepen, Acta Metall. 25, 407 (1977).

[25] M. Bletry, P. Guyot, J. J. Blandin, J. L. Soubeyroux, Acta Mater. 54, 1257 (2006).

[26] V. Trappe, V. Prasad, L. Cipelletti, P. N. Segre, & D. A. Weitz, Nature 411, 772 (2001).

[27] Z. Zhang, N. Xu, D. T. N. Chen, P. Yunker, A. M. Alsayed, K. B. Aptowicz, P. Habdas, A. J. Liu, S. R. Nagel, & A. G. Yodh, Nature 459, 230 (2009)

[28] P. G. Debenedetti, and F. H. Stillinger, Nature **410**, 259 (2001).

[29] W. L. Johnson, J. Lu, M.D. Demetriou, Intermetallics 10, 1039 (2002).

[30] X. L. Fu, Y. Li., C. A. Schuh, J. Mater. Res. 22, 1564 (2007).

Table 1. The values of $T_g$, shear modulus $G$, the compression yield strength $\sigma_y$, the maximum flow stress $\sigma_{max}$ and steady flow stress $\sigma_{flow}$ at the critical strain rate, the experimental critical strain rate $\dot{\gamma}_g$ and the calculated critical strain rate $\dot{\gamma}_g^{cal}$ of MGs.

| MGs | $Sr_{20}Ca_{20}Yb_{20}Mg_{20}Zn_{20}$ | $Sr_{20}Ca_{20}Yb_{20}Mg_{20}Zn_{10}Cu_{10}$ | $Sr_{20}Ca_{20}Yb_{20}(Li_{0.55}Mg_{0.45})_{20}Zn_{20}$ |
|---|---|---|---|
| $T_g$ (K) | 353 | 351 | 319 |
| $G$ (MPa) | 8.89 | 9.47 | 6.28 |
| $\sigma_y$ (MPa) | 452.1 | 433.6 | 375.1 |
| $\sigma_{max}$ (MPa) | 473.6 | 450.3 | 411.9 |
| $\sigma_{flow}$ (MPa) | 366.9 | 416.2 | 249.3 |
| $\dot{\gamma}_g (s^{-1})$ | $1\times10^{-6}$ | $5\times10^{-7}$ | $2\times10^{-4}$ |
| $\dot{\gamma}_g^{cal} (s^{-1})$ | $1.51\times10^{-6}$ | $1.45\times10^{-6}$ | $1.25\times10^{-4}$ |

**Figure captions**

Fig. 1. The compressive stress–strain curves at different strain rates of (a) $Sr_{20}Ca_{20}Yb_{20}Mg_{20}Zn_{20}$ and (b) $Sr_{20}Ca_{20}Yb_{20}(Li_{0.55}Mg_{0.45})_{20}Zn_{20}$ MGs. (c) the pictures of brittle behavior above $\dot{\gamma}_g$ and (d) homogeneous deformation behaviors (about 50% deformed) below the critical strain rate for $Sr_{20}Ca_{20}Yb_{20}Mg_{20}Zn_{20}$ MG.

Fig. 2. (a) Macroscopic and (b) microscopic SEM image of about 50% heavily deformed $Zn_{20}Ca_{20}Sr_{20}Yb_{20}(Li_{0.55}Mg_{0.45})_{20}$ MG. No shear bands and micro cracks can be found indicating the homogeneous deformation of the glass.

Fig. 3. The diagram of brittle to homogeneous flow translation induced by strain rate in $Sr_{20}Ca_{20}Yb_{20}(Li_{0.55}Mg_{0.45})_{20}Zn_{20}$ MG, plotting the maximum yield stress vs. the applied strain rate during compression. A well-defined transition from supercooled liquid state to glassy state can be found as $\dot{\gamma}$ reached a constant value. And a critical strain rate $\dot{\gamma}_g$, which is equivalent to $T_g$, exists in the strain rate induced glass transition. The curves are guide to the eyes.

Fig. 4. (a) The maximum yield stress vs. the inverse strain rate in the $Sr_{20}Ca_{20}Yb_{20}(Li_{0.55}Mg_{0.45})_{20}Zn_{20}$ MG. (b) The temperature vs. $\dot{\gamma}_g$ for $Sr_{20}Ca_{20}Yb_{20}Mg_{20}Zn_{10}Cu_{10}$ MG. The curves are guide to the eyes.

Fig. 5. The schematic 3D glass transition phase diagram involved in observation time $t$, temperature $T$, and stress $\sigma$ for MGs. Data from the strain rate induced glass transition in SrCaYbMg(Li)Zn(Cu) MGs are used to construct the phase diagram. This phase diagram is a compendium of various MG systems. The glass region, near the origin, is enclosed by the depicted concave surfaces.

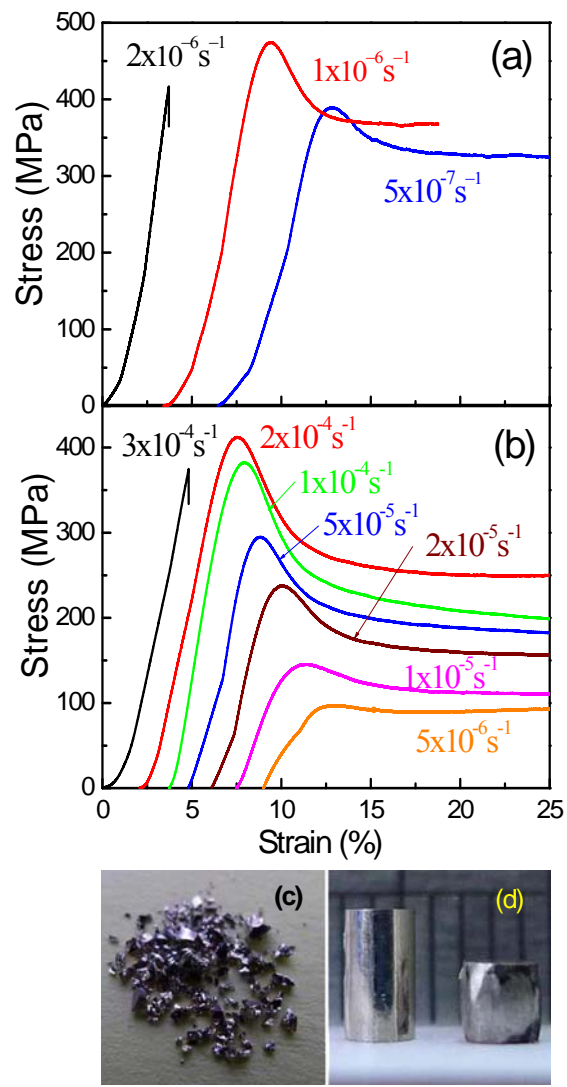

Figure 1, Gao *et al.*

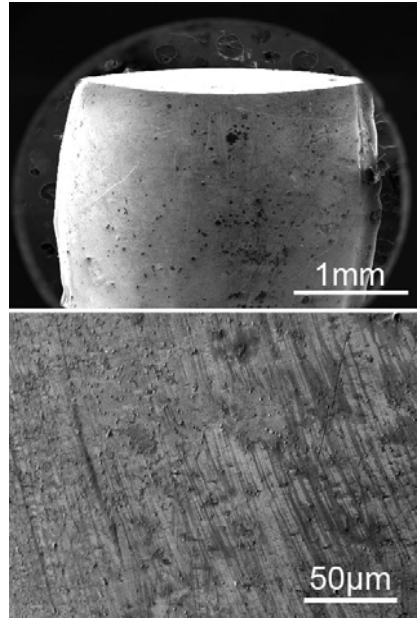

Figure 2, Gao *et al.*

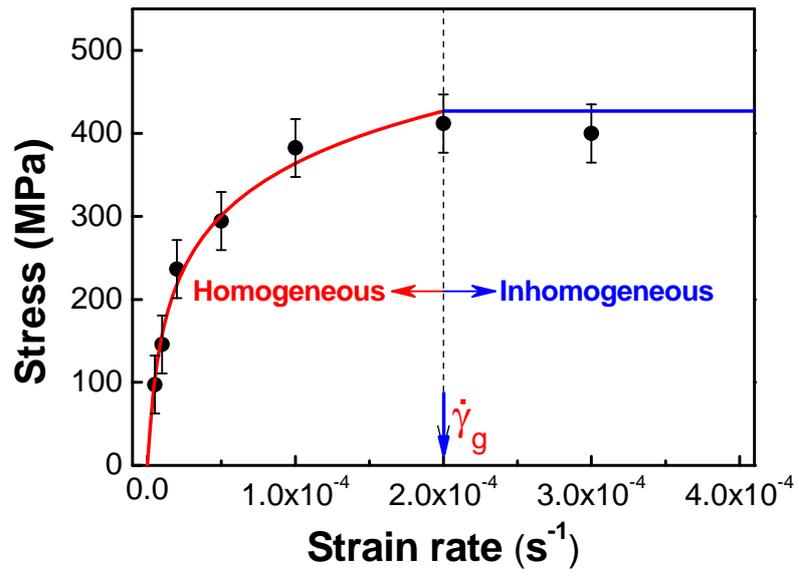

Figure 3, Gao *et al.*

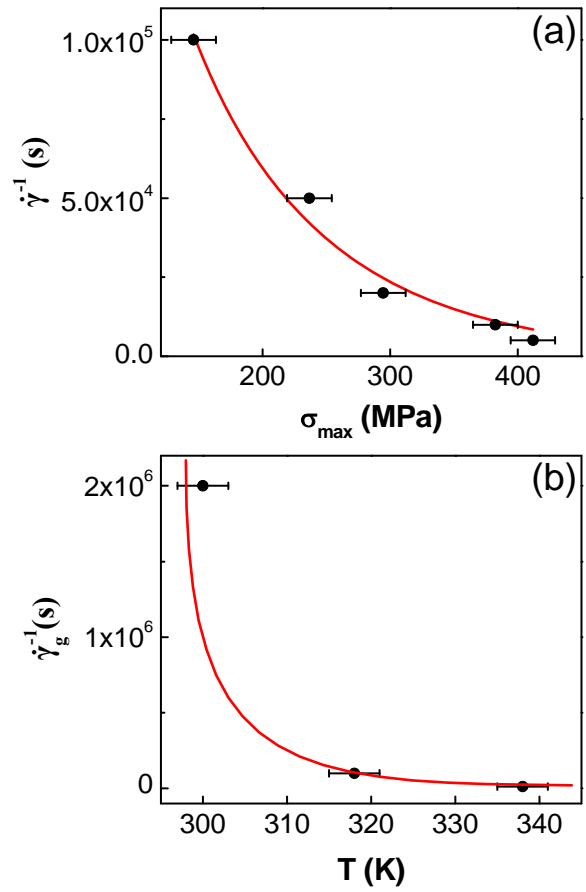

Figure 4, Gao *et al.*

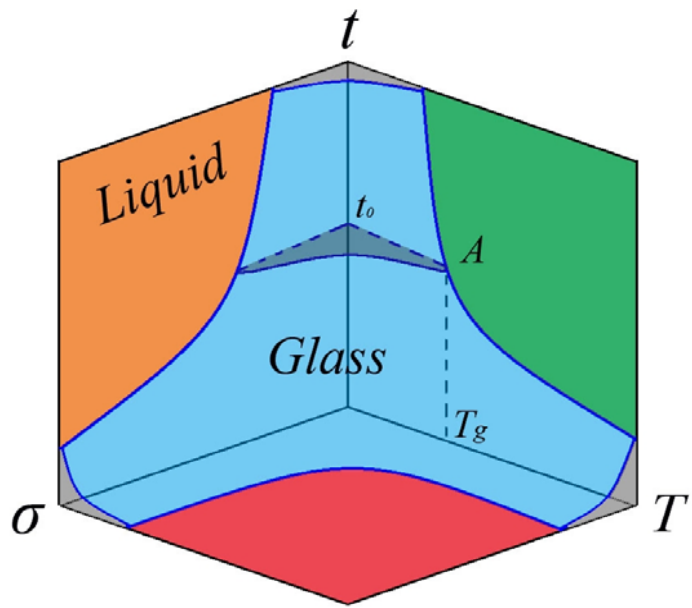

Figure 5, Gao *et al.*